\documentclass{jfm}
\usepackage{natbib,amsmath,amssymb}
\usepackage{psfig,epsfig,epsf}
\usepackage{dcolumn}

\newcommand{\beq}{\begin{equation}}  
\newcommand{\eeq}{\end{equation}}  
\newcommand{\bea}{\begin{eqnarray}}  
\newcommand{\eea}{\end{eqnarray}}  
\title[Transition to turbulence in pipe flow]{Sensitive dependence on initial
conditions in transition to turbulence in pipe flow}
\author[Holger Faisst and Bruno Eckhardt]%
{H\ls O\ls L\ls G\ls E\ls R\quad   F\ls A\ls I\ls S\ls S\ls T\ns \and
 B\ls R\ls U\ls N\ls O\quad   E\ls C\ls K\ls H\ls A\ls R\ls D\ls T
}
\affiliation{Fachbereich Physik, Philipps-Universit\"at Marburg,
D-35032 Marburg, Germany}
\pubyear{2004}
\volume{999}
\pagerange{1--10}
\date{??}
\begin{document}
\maketitle
\begin{abstract}
The experiments 
by Darbyshire and Mullin (J. Fluid Mech. {\bf 289}, 83 (1995))
on the transition to turbulence in pipe flow show that
there is no sharp border between initial conditions that
trigger turbulence and those that do not. We here relate this
behaviour to the possibility that the transition to turbulence
is connected with the formation of a chaotic saddle in the
phase space of the system. We quantify a sensitive dependence
on initial conditions and find in a statistical analysis that
in the transition region the distribution of turbulent lifetimes 
follows an exponential law. The characteristic mean lifetime 
of the distribution increases rapidly with Reynolds number and becomes
inaccessibly large for Reynolds numbers exceeding about 2200.
Suitable experiments to further probe this concept are proposed.
\end{abstract}     


  
\section{Introduction}

The transition to turbulence in pipe flow has been
the subject of many investigations since the first
documentation of its phenomenology in~\cite[]{reynolds1883}. 
Reynolds already noticed that there was no sharp transition,
that the transition could be delayed to very
high Reynolds number when perturbations to the
laminar profile were carefully avoided and
that sometimes turbulence appeared without
clear evidence for a perturbation that may have
triggered it. His findings have been confirmed and
expanded on in many subsequent experiments, 
e.g.~\cite[]{wygnanski73,wygnanski75,rubin80,eggels94,ma99,darbyshire95}.
On the theoretical side, no linear instability could
be found (see~\cite{schmid94,meseguer03} and 
references therein)
but analysis of the consequences of the non-normality 
of the linearized problem has led to the identification
of efficient amplification mechanisms that can
explain the sensitivity to small perturbations
of the laminar 
profile~\cite[]{boberg88,schmid94,trefethen00,trefethen93,
grossmann00,hof03}. 
On the other hand the simulations in~\cite[]{brosa88} and
the experiments in~\cite[]{darbyshire95} 
suggest that even if the perturbations are large enough to
trigger turbulence, the flow can re-laminarize
without any previous indication. Similar behaviour
has been found in plane Couette 
flow~\cite[]{bruno97,arminEPL,bottin98B,faisst00,bruno02}.

Decay of the turbulent flow must follow from the dynamics of 
the fully developed 3-d turbulent state and cannot be explained 
by linearization around the laminar flow. The dynamical system 
concept compatible with such a behaviour is that of a chaotic 
saddle, a transient object with chaotic dynamics~\cite[]{tel91,ott93}. 
The simplest example of a 
chaotic saddle arises for a particle in a box with a tiny hole: 
the dynamics inside the box can be chaotic, as measured by 
positive Lyapunov exponents, but it is transient and ends 
when the particle leaves through the hole. The escape through the
hole is a global event, and its rate is not related to 
the Lyapunov exponent. The analogy then is that the turbulent
state is motion in the box, and escape from the box is 
related to relaminarization. The characteristic signatures 
of a chaotic saddle are: 
(a) a sensitive dependence of lifetimes on initial conditions, 
(b) an exponential distribution of lifetimes for initial conditions around 
the chaotic saddle, 
(c) a positive Lyapunov exponent in the chaotic phase, and
(d) an independence of the variations of Lyapunov exponents
and the escape rates with parameters. 
We will show here that the transition to turbulence in pipe flow 
shows all these characteristics and is thus compatible with the 
formation of a chaotic saddle.

Our analysis is based on numerical simulations of the time evolution
of perturbations in circular pipe flow with periodic
boundary conditions in the downstream direction. The axial
periodicity length will be too short to simulate localized turbulent 
structures, such as puffs and slugs~\cite[]{wygnanski73,wygnanski75}, 
but long enough to capture the local turbulent structures. Following
the simulations of~\cite[]{eggels94} we take a length $10R$, 
with $R$ the pipe radius. The turbulence will then typically fill 
the entire volume and we do not have to consider the advection of
the turbulent state by the mean profile: lifetimes can be defined 
locally. The Reynolds number $Re$ is based on 
the mean streamwise flow velocity $U$ and the pipe diameter $2R$,
\beq
Re =  2R U /\nu\,.
\eeq
and the units of time are $2R/U$. 
As in the experiments~\cite[]{darbyshire95} we keep the volume flux
constant in time: this simplifies
the analysis and prevents decay due to a reduction of flux
that could occur in pressure driven situations when
the flow becomes turbulent.

\section{Numerical considerations}
We use cylindrical coordinates and employ a pseudo-spectral 
method with Fourier modes in the periodic azimuthal and 
downstream direction and Legendre collocation radially~\cite[]{canuto88}. 
Various linear constraints on the velocity field are treated 
together by the method of Lagrange multipliers: 
the rigid boundary condition, the solenoidality, 
and the analyticity 
in the neighbourhood of the coordinate singularity at the 
center line~\cite[]{priymak98}.  
We verified our numerical scheme in various ways. 
For the linearized dynamics we reproduced the literature data on 
the eigenvalue spectrum of the linearized Navier-Stokes operator
with full spectral precision~\cite[]{schmid94}.
For the non-normal linear dynamics and nonlinear 2D dynamics 
we reproduced Zikanov's results~\cite[]{zikanov96}. 
At $Re=5000$ a long, fully nonlinear 3D turbulent trajectory 
was analysed and its statistical properties agreed with previous 
numerical and experimental results~\cite[]{eggels94,quadrio00}. 

The results presented here are obtained with
azimuthal and axial resolution of $|n/17|+|m/15|<1$, 
where $n$ and $m$ are the azimuthal and streamwise wavenumbers, 
respectively, and $50$ Legendre polynomials radially.
This resolution is not the maximal that could be integrated 
for a single run but reflects a compromise between the mutually 
exclusive requirements of maximal resolution, maximal integration 
time for a single run and a large number of sample runs for the 
statistical evaluations. It is justified by comparisons with 
lower and higher resolutions which show no significant
differences for the range of Reynolds numbers 
$Re<2300$ to which we restrict our computations. 
The maximal integration time after which we terminate a 
turbulent trajectory is $2000$, sometimes $3000$ natural time 
units. This by far exceeds the values accessible in the longest 
currently available experimental set up~\cite[]{ma99,hof03}.
For the statistical analysis and the calculations of Lyapunov exponents 
more than 1000 runs were needed, adding up to several years
of CPU time on a single 2.2 GHz Pentium4/Xeon processor.

The initial conditions for each run are the parabolic 
profile ${\bf u}_{HP}$ to which a three-dimensional 
perturbation ${\bf u}_p$ is added. 
The specific form of ${\bf u}_p$ should not matter,
as long as it triggers a transition to turbulence, since
the dynamics of the turbulent state has positive Lyapunov 
exponents and leads to a quick elimination of any details 
and memory of the initial conditions.
Furthermore, experiments with different kinds of wall-normal
or azimuthal jets or other perturbations lead to very similar
results~\cite[]{darbyshire95}, 
so that we can safely assume that there is only one turbulent
state to which all initial conditions are attracted.
Therefore, we take as initial condition an uncorrelated random 
superposition of all available spectral modes.
Changes in the initial conditions are then limited to 
variations in the amplitude of the perturbation, not
in the form, i.e., throughout this paper 
we scan the behaviour along a 
one-dimensional subset in the space of all velocity 
fields, ${\bf u}(0)={\bf u}_{HP} + a {\bf u}_p$. 

We define the amplitude $A$ of an initial disturbance as its 
kinetic energy in units of the energy of the laminar 
Hagen-Poiseuille profile ${\bf u}_{HP}$, 
\begin{equation}
A= a^2 \frac{\int dV {\bf u}_p^2}{\int dV {\bf u}_{HP}^2} \, .
\end{equation}

\begin{figure}
\begin{center}
\epsfig{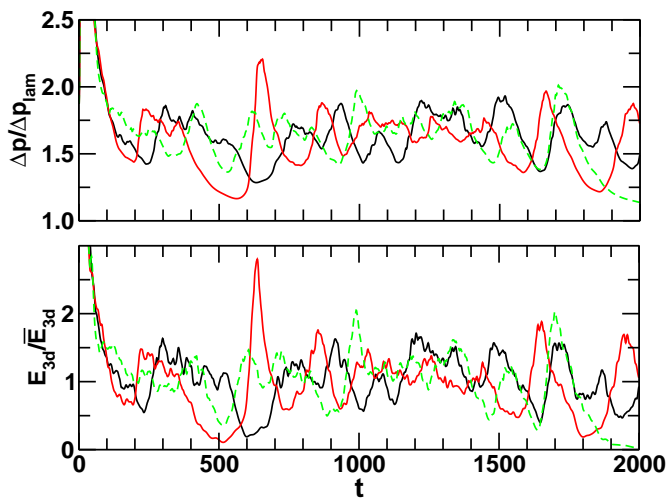}
\end{center}
\caption[]{
Traces of normalized pressure gradient (top) and energy 
content $E_{3d}$ (Eq.~\ref{e_3d}) (bottom) for  three slightly 
different initial conditions at $Re=2100$. The initial energies 
are chosen high above the turbulent mean and the trajectories 
require about $200$ time units to relax to the neighborhood of 
the turbulent mean. Pressure drop and energy content show 
correlated fluctuations. Note that within statistical 
fluctuations the averages of the turbulent dynamics are 
the same. This includes the dashed trajectory which happens to 
decay near $t=2000$.
The bottom frame shows the kinetic energy of the 
streamwise modulated part 
of the velocity field, $E_{3d}$, 
normalized by its turbulent mean
as a measure of the velocity fluctuations
\label{typical_trajects}
}
\end{figure}

The measures that we apply to the turbulent runs are the 
pressure gradient required to maintain a constant mean 
flux and the energy variable $E_{3d}$, the kinetic energy 
in the streamwise modulated part of the velocity field,
\beq
E_{3d} = 
\sum_n\sum_{m\neq 0} \int_0^R |{\bf u}_{n,m}(r)|^2\,r\,dr  
\, .
\label{e_3d}
\eeq
The significance of $E_{3d}$ is that if it becomes too small, 
then the flow  is too close to an axially translation invariant 
flow field and will eventually decay~\cite[]{zikanov96}. We 
therefore terminate integration if it drops below a threshold 
of $10^{-4}$.

\section{Turbulent lifetimes}
Typical trajectories are shown in Fig.~\ref{typical_trajects}. 
Within about $150$ time units they relax towards the turbulent state.
The dashed trace belongs to an initial condition for
which the lifetime of the turbulent state is about $2000$: there is 
no indication of the decay until the energy $E_{3d}$ drops so low 
that turbulence cannot recover. 

\begin{figure}
\begin{center}
\epsfig{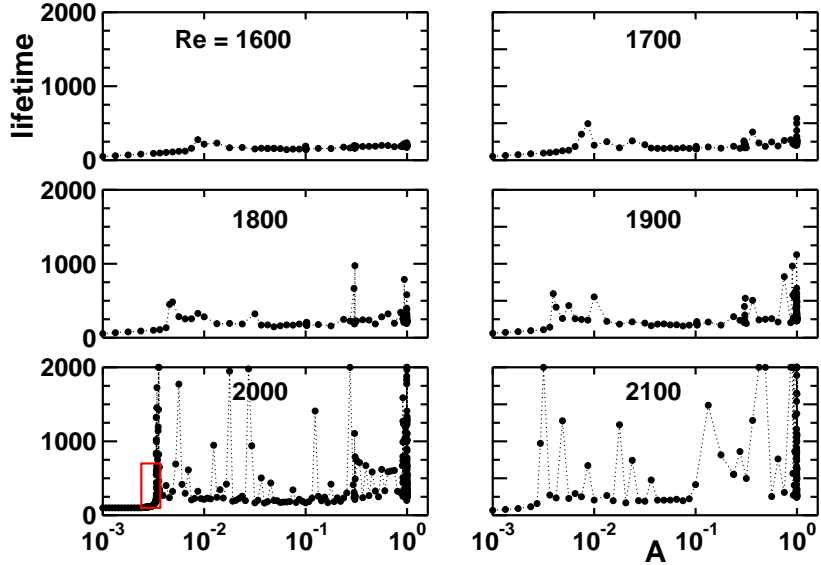}
\end{center}
\caption[]{
Turbulent lifetime vs.~perturbation amplitude for different 
Reynolds numbers.  For all Reynolds numbers a smooth
region with short-lived states can be observed for small 
amplitudes. For larger Reynolds numbers this is followed 
by a ragged region of highly fluctuating lifetimes. 
The threshold amplitude that divides the two regions decreases 
with Reynolds number. The little rectangle at amplitudes of 
about $0.03$ at $Re=2000$ is further magnified in 
Fig.~\ref{lt_2k}
\label{lifetime_slices}
}
\end{figure}

Scanning the lifetimes of turbulent states as a function of initial
amplitude for different Reynolds number we obtain the results shown 
in Fig.~\ref{lifetime_slices}. For sufficiently
small amplitudes all states decay and the lifetimes are short.
Beginning with a Reynolds number of $1800$ sharp peaks with very long
lifetimes appear. Beyond $Re = 2000$ several initial conditions
reach lifetimes up to the integration cut-off of $t_{max}=2000$.
Moreover, the variations in lifetimes between neighboring
sampling points in amplitude increase. 

\begin{figure}
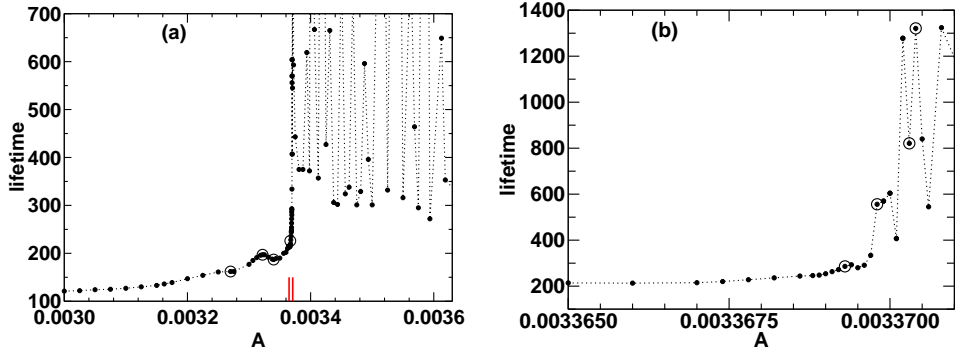

\begin{center}
~\hfill
\epsfig{file=pipe_trans_fig/pipe_trans3a_NEW.eps,width=.45\textwidth,clip=}
\hfill
\epsfig{file=pipe_trans_fig/pipe_trans3b_NEW.eps,width=.45\textwidth,clip=}
\hfill~
\end{center}
\caption[]{Successive magnifications of the lifetime variations with
amplitude in the transition region for $Re=2000$. The left panel
is a magnification of the box indicated in Fig.~\ref{lifetime_slices},
the right one a magnification of the interval indicated in the left panel.
Note the increased ordinate scale in the right panel.
\label{lt_2k}
}
\end{figure}

The transition between the short lifetimes for the initial conditions
that decay quickly towards the laminar state and the longer ones
for trajectories that show some turbulent behaviour is a very rapid
one. Successive magnifications of a small region in amplitude for
$Re=2000$ are shown in Fig.~\ref{lt_2k}.
The increase is not monotonic, with modulations and structures
superimposed.

\begin{figure}
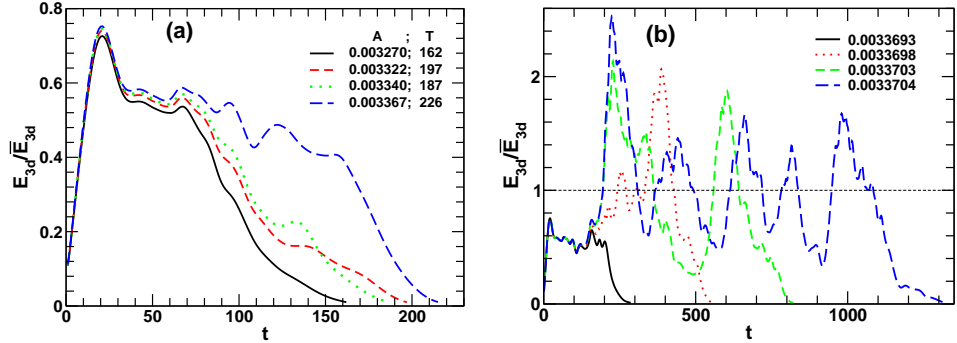

\begin{center}
~\hfill
\epsfig{file=pipe_trans_fig/pipe_trans4a.eps,width=.45\textwidth,height=.335\textwidth,clip=}
\hfill
\epsfig{file=pipe_trans_fig/pipe_trans4b.eps,width=.45\textwidth,height=.33\textwidth,clip=}
\hfill~
\end{center}
\caption[]{
Energy traces for trajectories in small intervals of amplitude 
close to the transition region at $Re=2000$, marked by open circles 
in the corresponding panel in  Fig.~\ref{lt_2k}. 
The small differences in initial conditions are amplified and
lead to noticeable differences in their evolution. For panel (b) 
the lifetimes are larger and the dynamics shows several large 
amplitude oscillations connected with the turbulent regeneration 
dynamics.
\label{nonmono}
}
\end{figure}

Fig.~\ref{nonmono}
compares the energy traces for four initial conditions 
in two ranges of amplitude. 
Fig.~\ref{nonmono}a for the
smaller mean amplitude shows that in an interval 
of width about $3\%$ of the mean amplitude the lifetimes 
increase by about $40\%$. 
At a slightly higher amplitude Fig.~\ref{nonmono}b shows
that in an interval of relative width $3\cdot10^{-4}$
the lifetimes increase by a factor of almost $4$!

\begin{figure}
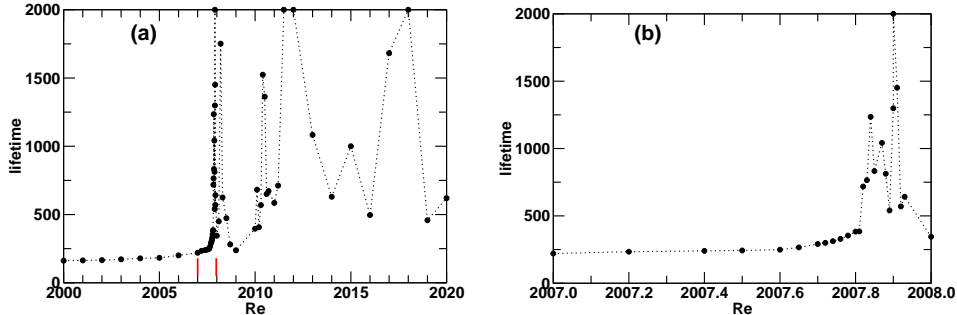

\begin{center}
~\hfill
\epsfig{file=pipe_trans_fig/pipe_trans5a.eps,width=.45\textwidth,clip=}
\hfill
\epsfig{file=pipe_trans_fig/pipe_trans5b.eps,width=.45\textwidth,clip=}
\hfill~
\end{center}
\caption[]{
Variations of turbulent lifetimes with Reynolds number for a fixed initial 
disturbance amplitude $A=0.00327$.
The parameter range marked in (a) by the shaded interval 
near $Re=2007$ is magnified in (b).
\label{cut_re}
}
\end{figure}

The strong fluctuations in lifetimes are not limited to 
variations with amplitude alone, they also occur
when changing the Reynolds number for fixed amplitude, 
as shown in Fig.~\ref{cut_re}. Such fractal behaviour
in the variations of lifetimes under parameter variation
has previously been observed in plane Couette flow~\cite[]{bruno97},
Taylor-Couette flow for large radii~\cite[]{faisst00,bruno02},
and in models with a sufficient number of degrees of freedom~\cite[]{eckhardt99}.

\section{Lyapunov exponents}
The sensitive dependence on variations of initial conditions
can also be quantified in terms of Lyapunov 
exponents.
For a trajectory ${\bf u}(t)$ and deviations $\delta{\bf u}(t)$ 
the largest Lyapunov exponent is defined as
\begin{equation}
\lambda=\lim_{||\delta u(0)||\rightarrow 0}\,\, \lim_{T\rightarrow\infty}\, 
\frac{1}{T} \log \frac{||\delta u(T)||}{||\delta u(0)||}\, , 
\end{equation}
where $||\cdot||$ denotes the Euclidian norm. 
Instead of a single $\lambda$ from an infinite time interval we calculate ensembles of 
finite-time Lyapunov exponents, extracted
from the integration of a small deviation
from the reference trajectory.
The separation was measured after time intervals of $T=200$ units,
and then rescaled to $||\delta u||=10^{-9}$.  
An initial time interval of $200$ units was omitted because
of the transient relaxation of the initial conditions onto the
turbulent state. Similarly, the last $200$ time units were omitted
to avoid the decay to the laminar state. 
The infinite-time Lyapunov exponent was then approximated as the average of 
the finite-time Lyapunov exponents. 

\begin{figure}
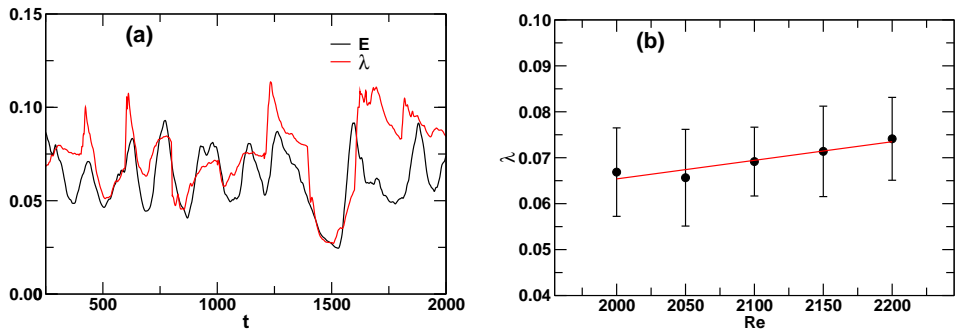

\begin{center}
~\hfill
\epsfig{file=pipe_trans_fig/pipe_trans6a.eps,width=.45\textwidth,clip=}
\hfill
\epsfig{file=pipe_trans_fig/pipe_trans6b.eps,width=.45\textwidth,clip=}
\hfill~
\end{center}
\caption[]{
\label{en_ly}
Lyapunov exponents in pipe flow. (a) Local Lyapunov exponent 
for a Reynolds number of $Re=2150$. The 
fluctuations in the local Lyapunov exponent correlate strongly with the 
energy content $E$ of the disturbance 
(in units of the energy of the laminar profile).
(b) Largest Lyapunov exponent of the turbulent state with
an error estimate based on the root mean square deviations 
of the ensemble of finite-time Lyapunov exponents. The line
indicates a linear fit with slope $d\lambda/d Re\approx 4\times 10^{-5}$.}
\end{figure}

The instantaneous changes in the separation between neighbouring
trajectories can be used to define local Lyapunov exponents
\cite[]{eckhardt93}. They correlate strongly with large
energy fluctuations, 
see Fig.~\ref{en_ly}a. When new large 
scale structures are generated the energy grows strongly and the Lyapunov 
exponent increases.  Towards the end of a nonlinear regeneration cycle 
the energy goes down and the Lyapunov exponent decreases as well. 
Therefore, the fluctuations in the exponent are large 
and averages over at least $10^4$ time units are needed
for the determination of reliable numbers. Obviously, this is very
difficult to achieve for Reynolds numbers below $Re=2000$, 
where only very few trajectories stay turbulent 
for sufficiently long times.

The largest Lyapunov exponent is shown in Fig.~\ref{en_ly}b.
Its typical value is about $6.5\times 10^{-2}$ at 
Reynolds numbers around $2000$. 
For predictability this Lyapunov exponent implies
that an uncertainty doubles after a time of 
$(\ln 2)/\lambda \approx 10$ units. 
Over a time interval of $200$ units a perturbation grows 
$10^6$-fold! 
Note that $200$ units is about the separation between
two oscillations in the mean energy in 
Figs.~\ref{typical_trajects} and~\ref{en_ly}a and 
is the typical duration of the regeneration cycle 
proposed in~\cite[]{hamilton95,waleffe95,waleffe97}. 
Thus, the Lyapunov exponent is small in absolute value, 
but fairly large on the intrinsic time scales 
of the turbulent dynamics.

The extreme sensitivity of turbulent lifetimes to
variations in initial conditions will make it next
to impossible to prepare experimental perturbations
sufficiently accurately to reproduce a run. A
statistical analysis of, e.g., the distribution of 
lifetimes obtained by collecting data for several 
nearby initial conditions, should be more
reliable. We therefore study $P(t)$, the
probability that a flow started with some
initial condition will still be turbulent after a 
time $t$. 
A related kind of statistics was proposed in~\cite[]{darbyshire95}: 
they analyzed the fraction $p$ of initial
conditions that remained turbulent over the time 
$T$ it took for the perturbation to transit the
distance between perturbation and detection. 
(Recall that we do not need to take this advection into
account since our periodicity length is so short that the
turbulence fills the entire volume.)
In terms of our $P(t)$, this probability is
$p=P(T)$. However, because of the unknown 
relation between the initial relaxation times
in our simulations and the experiment we cannot
compare values.

\begin{figure}
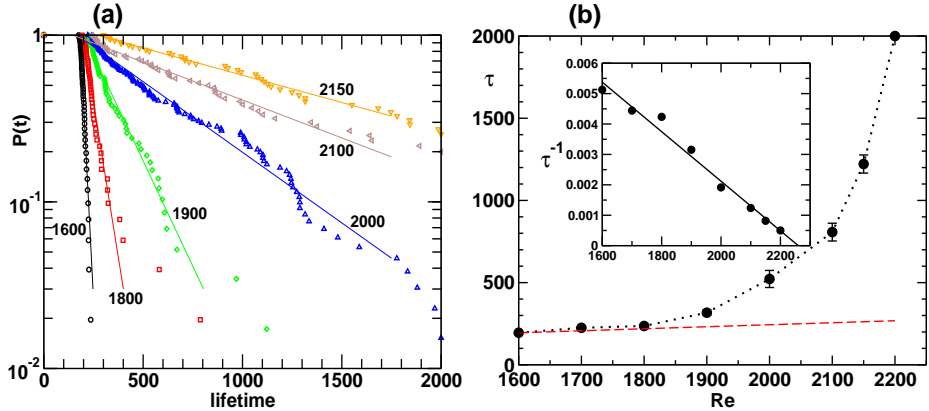

\begin{center}
\epsfig{file=pipe_trans_fig/pipe_trans7a.eps,width=.45\textwidth,height=.40\textwidth,clip=}
\epsfig{file=pipe_trans_fig/pipe_trans7b.eps,width=.45\textwidth,height=.40\textwidth,clip=}
\end{center}
\caption[]{Turbulent lifetimes for different transitional Reynolds numbers.
(a) Probability $P(t)$ for a single trajectory to still be turbulent after 
a time $t$ for six Reynolds numbers as indicated.
Between $50$ and $100$ trajectories have
been evaluated per Reynolds number. 
The distributions are well described by exponential distributions shown as
straight lines.
(b) Median $\tau$ of the turbulent lifetimes as a function of 
Reynolds number. The median and the fluctuations increase rapidly 
with Reynolds number until the median reaches 
the cut-off lifetime of $2000$ at $Re=2200$. 
The error bars indicate the statistical uncertainty of the median.
The straight dashed line shows the linear increase in lifetime 
expected due to purely non-normal linear dynamics.
The inset shows the inverse median lifetime vs.~$Re$ and a linear 
fit, corresponding to a law $\tau(Re)\propto (Re_c-Re)^{-1}$, with 
$Re_c \approx 2250$.
\label{life_median}
\label{life_dist}
}
\end{figure}

For the lifetime statistics we used more than $50$ 
initial amplitudes each for eight Reynolds numbers in the range
from $1600$ to $2200$. The amplitudes were fairly 
high, $\approx 1$, in order to assure transition 
to turbulence at least for a short time interval.  
The results for different Reynolds numbers
are shown in Fig.~\ref{life_dist}. 

The data support an exponential decay $P(t) \sim \exp(-\epsilon t)$ 
for large times, with $\epsilon$ the rate of escape 
from the turbulent state.
An exponential distribution of turbulent lifetimes 
is a characteristic signature for escape from a 
chaotic saddle~\cite[]{kadanoff84,tel91,ott93}.  
They have previously 
been seen in experiments~\cite[]{bottin98A,bottin98B} and 
numerical studies on plane Couette flow~\cite[]{schmiegel_phd}, 
as well as in Taylor-Couette flow~\cite[]{bruno02}.
They seem to be a generic feature of the transition in shear 
flows that are not dominated by linear instabilities.

Another measure that can be extracted from the lifetime
distributions is the median, $\tau = (\ln 2)/\epsilon$.
The median is the time up to which half the states have decayed.
It is also well suited for numerical studies since it is
not affected by the cut-off in integration time.
The median increases rapidly with Reynolds number as shown in
Fig.~\ref{life_median}b.
Below $Re=1800$ the increase is mainly due to non-normal 
transient linear dynamics~\cite[]{grossmann00}. 
When the Reynolds number is 
increased above $Re=2000$ the median of the turbulent 
lifetimes as well as the fluctuations rise rapidly 
until the median reaches the cut-off lifetime of $2000$ 
at $Re \approx 2200$.
The inset in Fig.~\ref{life_median} shows the inverse median 
lifetime vs.~$Re$ and among the possible fits for these 
data the most satisfactory ones supports 
a divergence like $(Re_c-Re)^{-1}$, with $Re_c \approx 2250$.
Such a divergence would be connected with a transition
from the chaotic saddle for lower $Re$ to an attractor for
larger $Re$. However, it is known from other models that
the lifetimes can increase rapidly without a true singularity~\cite[]{tel91,crutchfield}.
The question of whether we will arrive
at a turbulent attractor cannot be answered here. 
But from the rapid increase it is clear that it will become
an attractor for all practical purposes: for a setup with
length $1000$ diameters the median will exceed this
value for Reynolds number about $2130$. For this
observation time the fraction of states with shorter
lifetimes drops from $80\%$ of all initial conditions for
$Re=2000$ to $40\%$ for $Re=2150$ and to zero for Re=$2200$.
These numbers are obtained from the numerical simulations and 
include the initial transient; numbers in other simulations
and experiments may be different, but the rapid increase
will remain the same.

\section{Conclusions}
An exponential distribution of lifetimes can be obtained
for a constant probability of decaying. In view
of the internal dynamics and the characteristic times
connected with it, one can interpret this as the 
probability of decaying towards the laminar state at
the end of each turbulent regeneration cycle.
The conclusion from the exponential distribution of 
lifetimes is that this probability remains constant 
during the evolution, independent of the `age' of the trajectory.

It is worthwhile pointing out that the strong variation of the
median lifetimes with $Re$ is not reflected in the Lyapunov exponent, which
increases only linearly with $Re$: this shows that the chaotic dynamics
on the turbulent saddle and the escape from it are two different
processes with independent characteristics.

There are two predictions that should be accessible to experimental
investigation: the behaviour near the boundary between laminar and
turbulent and the exponential distribution of lifetimes.
The increase in lifetime for increasing perturbation amplitude
can be analyzed with a series of sensors along the pipe.
The oscillations in Fig.~\ref{en_ly}a have a period of about $200$ time
units, which translates into $200$ diameters. 
Thus, if a perturbation can be localized within 
a few radii, detectors that are further and further downstream
might be able to reproduce some of the oscillations
in Figure 4 as the perturbation amplitude is increased. 
Similarly, it might be possible to experimentally detect
the Lyapunov exponents in the flow: 
with a Lyapunov exponent of about $0.07$, 
a $5\%$ uncertainty in the preparation of initial conditions
will increase to $100\%$ in a time of about $40$ units, i.e. $40$ diameters.  
Thus, 
placing detectors at spacings of several diameters apart
should allow a determination of the separation between
flow states starting from similar initial conditions.

The second measurement is more easily performed: repeated runs
with similar perturbations give the probability of  
finding a turbulent state that exists at least up to time $t$. 
The time can be varied again by placing detectors at different
locations along the pipe. As explained, the exponential
distribution of lifetimes is a
property of the turbulent state in the transition region and will not
depend critically on the type of initial conditions used.
Exact reproduction of an initial condition is hence not 
essential here since all initial
conditions relax to the same turbulent state. The median lifetime
increases rapidly with Reynolds number, from about $220$ at $Re=1800$
to $750$ at $Re=2100$. Given the experimental limitations on the length
of the pipe this region looks like a promising
range for experiments. The mechanism studied here, the formation of
a chaotic saddle, is fairly independent of the boundary conditions, but 
the quantitative characteristics may depend on it. Since the experiments
will not have periodic boundary conditions in downstream direction
and since the turbulence may be localized in puffs or slugs it will be
interesting to see whether this will affect the lifetimes
or its distribution.

A detailed analysis of the distribution of lifetimes is a task which  
is still much better suited for experiments than for numerics: it is not
necessary to prepare the initial disturbance with highest precision, and
the experimental observation time is of the order of $1$ minute for an
experimental run, followed by about half an hour for the water to return 
to rest in the head-tank. On the other hand, a single trajectory in our 
computation takes up to $10$ days.

The picture that emerges from these data is that the transition to 
turbulence in pipe flow is connected with the
formation of a chaotic saddle. The turbulent dynamics is chaotic,
the lifetimes are exponentially distributed and the transition
depends sensitively on the initial state. The same
characteristics have been found in other shear flows without
a linear instability as well~\cite[]{bruno97,bruno02,bottin98B,faisst00}. 
The recent discovery of travelling waves in pipe flow for
Reynolds numbers as low as $1250$~\cite[]{FE_TW03,TW_Bristol}
strengthens this picture by providing some of the
states around which the network of homoclinic
and heteroclinic connections that carry the chaotic
saddle may form.\\

Support by the German Science Foundation is greatfully acknowledged.

\bibliography{pipe_trans_jfm}
\bibliographystyle{jfm}
\end{document}